# The Nature of the Compact HII Region Sh 2-89 and its Stellar Content

G. Ramos-Larios, J.P. Phillips, J. A. Pérez-Grana

Instituto de Astronomía y Meteorología, Av. Vallarta No. 2602, Col. Arcos Vallarta, C.P. 44130 Guadalajara, Jalisco, México   e-mail : jpp@astro.iam.udg.mx

**Abstract**

We present an analysis of the structure and properties of the compact HII region Sh 2-89, and certain of the young stellar objects (YSOs) within this regime, using mid-infrared (MIR) mapping derived from the *Spitzer Space Telescope* (*SST*) and visual slit spectroscopy of the inner regions of the source. We show that the region has a bipolar structure, and contains a variety of Class I and II YSOs. Much of the MIR emission appears to be dominated by PAH emission bands, which cause strong increases in flux in the 5.8 and 8.0 μm photometric channels, whilst the variation of Hα, [NII] λ 6583 Å, [SII] λλ 6716+6731 Å, and MIR emission profiles confirms the presence of complex ionisation fronts, and ionisation stratification. One of the sources in this region has been previously identified as a likely symbiotic star. We show however that whilst it contains TiO absorption bands, it shows little evidence for higher excitation HeII λ4686 Å or [OIII] λ5007 Å emission. Such a result does not rule out the possibility of the source being a symbiotic star, although we note that the detection limits on these lines, and the observed visual/infrared continuum would make it difficult to fit it into either the D- or S-type symbiotic classifications. An analysis and modelling of the infrared continuum, on the other hand, together with a consideration of the associated HII region, suggests that it is more likely to represent a Class II YSO in which most of the emission derives from a massive accretion disk. Where this is the case, then the star is likely to have a temperature $T_{EFF}$ ~$10^4 \rightarrow 2\ 10^4$ K, mass ~2→4 $M_\odot$, luminosity 0.5→3 $10^2$ $L_\odot$, and age in the range $10^6 \rightarrow 10^7$ years. It also appears to be associated with strong Hα emission which, where it derives from radiatively driven stellar/disk winds, and collimated or spherically symmetric mass outflows, would likely imply mass-loss rates of order dM/dt ~ 4.5 $10^{-8}$ - 4.5 $10^{-7}$ $M_\odot$ yr$^{-1}$ (depending upon the outflow velocity $V_W$). Such a mass-loss rate is entirely plausible for this category of source, although it is also noted that the Hα may derive from other mechanisms as well, including chromospheric activity, and the ionisation of circumstellar material – the result of interaction between local UV radiation fields and winds and the stellar accretion disk or envelope.

**Key Words:** (ISM:) HII regions --- ISM: jets and outflows --- (stars:) circumstellar matter --- stars: formation --- stars: mass-loss --- stars: pre-main-sequence--- stars: winds, outflows

## 1. Introduction

Galactic HII regions are known to possess a broad range of sizes, densities, morphologies and other physical characteristics. Thus for instance, the largest giant and "classical" HII regions have sizes > 10 pc, densities $\sim 10^2$ cm$^{-3}$, appear to be characterized by low density gradients (e.g. Phillips 2008), and may be excited be a variety of stars and star-formation centres. By contrast, the ultracompact HII regions (UC HII regions) have dimensions < 0.005 pc and densities > $10^4$ cm$^{-3}$ (see e.g Israel et al. 1973; Dreher & Welch 1981 for some of the early and defining papers on these sources, and surveys by Becker et al. 1994; Wood & Churchwell 1989; Garay et al. 1993; Kurtz et al. 1994; Churchwell 2002; and Miralles et al. 1994), whilst the so-called hyper-compact regions are an order of magnitude smaller yet again, and probably associated with the formation of massive stars in the cores of molecular clouds (see e.g. Kurtz 2000, and the compilation of these sources by Johnson et al. 1988). There is evidence that many of these more compact HII regions possess steep density gradients - comparable or greater, in some cases, to those observed in stellar winds (Kurtz & Franco 2002; Franco et al. 2000; De Pree et al. 1995; Jaffe & Martin-Pintado 1999; Phillips 2007).

The so-called compact HII regions, by contrast, have intermediate sizes $\sim$0.05-0.4 pc (see e.g. Mezger et al. 1967), with Sh 2-89 appearing to be typical of this category of source. Early observations summarized by Fich & Blitz (1984) placed it at a kinematic distance of D = 2.72 ± 1.36 kpc, and indicated a diameter $\sim$4 pc at the lower end of the classical HII size-range. Subsequent radio mapping by Reich et al. (1990) and Taylor et al (1996), however, the latter undertaken with beam sizes of 1'x1'csc $\delta$, suggested that the source was largely unresolved, and had dimensions of < 0.8 pc for the same kinematic distance.

The validity of this distance is somewhat in doubt, given that the HII and CO velocities of Crampton et al. (1978) and Blitz et al. (1982) appear to be at odds (respectively 12.7 km s$^{-1}$ and 25.6 km s$^{-1}$). The reddening distances of nearby stars yield a closely similar value (2.9 ± 0.2 kpc; Forbes 1989), however, and we shall assume that the distance to the source is close to this latter value. Allowing for the broad errors associated with these values, such a distance is comparable to those of other Sharpless sources in the region (87, 88, 90, 93), various large scale dust structures, and the Lynds clouds L810 and L812, implying that it may be physically related to the Vulpecula OB1 complex described by Turner (1986).

Although our information concerning this object is skeletal at best, it appears to represent a typical example of a small-scale star formation centre, conceivably triggered through the action of a supernova centred close to L 792 (Taylor 1986). It therefore came as a surprise to note that one of the putative symbiotic stars identified by Corradi et al. (2008) appears to be located very close to the centre of this particular source. If this identification is correct, and symbiotic and HII region are located at a similar distance from the Sun, then this might suggest that our identification of the source is somewhat askew. Perhaps the Sh 2-89 region represents a symbiotic nebula, one of the small group of extended structures associated with D-type systems.

It was therefore of interest to investigate the nature of the star identified by Corradi et al. (2008), and gain some clearer insight into the properties of the shell associated with Sh 2-89. We have attempted to undertake such a program though visual spectroscopy of the symbiotic candidate; through an analysis of the ionic ($H\alpha$, [SII] and [NII]) structure of the source; and through an investigation of the MIR characteristics of the Sh 2-89 shell.

We shall show, as a result of this work, that Sh 2-89 has a bilobal structure which is centred upon a relative bright Class I young stellar object (YSO), and that several other Class I–II YSOs are likely to litter this regime (see e.g. Andre (1994) and Barsony (1995) for an explanation of these terms). It will also be argued that although the candidate symbiotic star could correspond to an anomalous S- or D-type source, the environment in which it is located, and the observed visual-IR continuum are more consistent with it being a Class II YSO. Where this is the case, then radiative transfer modelling can be used to fit the observed spectral-energy distribution (SED), and we will use this to constrain various of the physical characteristics of the source. We shall finally show that the region is laced with multiple I-fronts and neutral regimes; regions which leave their distinctive (and differing) signatures upon the optical and infrared profiles.

## 2. Observations

Low dispersion optical spectra of Sh 2-89 were obtained on the 15$^{th}$ May 2009, using a Boller & Chivens spectrometer mounted at the prime focus of the 2.1 m telescope, Observatorio Astronomico Nacional (OAN-SPM), San Pedro Martir, Baja California, Mexico. A CCD SITe1 1024x1024 CCD was used as detector, in tandem with a grating having 300 l/mm set at a blaze of 5000 Å. The slit length was 5 arcmins, and the slit width was 180 $\mu$m ($\equiv$ 2.35 arcsec), whilst the plate and spectral scales were 1.05 arcsec pix$^{-1}$ and 2.36 Å pix$^{-1}$ respectively. The spectral resolution was 10.66 Å, the wavelength uncertainty of order ~4 Å, and the spectral range was 3500 Å→7200 Å - although the limits of this range have a lower S/N, and have been excluded from the present analysis. Three exposures of 15 min each were combined to give a total exposure time of 45 min, whilst mean air masses were low and of order 1.035. Finally the results were calibrated using the stars Hz 44 (for which the exposure time was 300 s, and air mass 1.17), and Feige 34 (exposure time 240 s, A.M. = 1.2). The fluxes for these stars are based on magnitudes deriving from Oke (1990), corrected for a systematic error of 0.04 mag (Colina & Bohlin 1994), and converted into fluxes $F_\nu$ (erg/cm$^2$/s/Å) using the relation of Hamuy (1992) (a description of this process is to be found at http://www.astrossp.unam.mx/estandar/standards/okestandards.html).

We have used these data to analyze the putative symbiotic star of Corradi et al. (2008) (see Sect. 1), the spectrum of which was contaminated with nebular components of emission. Care was taken to remove these components by determining a mean spectrum on either side of the star, and eliminating this from the results. We have also used the results to determine the physical properties of the HII region (Sect. 3.3), and variations in intensity for H$\alpha$, [NII] $\lambda$ 6583 Å, and [SII] $\lambda\lambda$ 6716+6731 Å.

Mid-infrared (MIR) mapping of Sh 2-89 was obtained using data deriving from the Galactic Legacy Infrared Midplane Survey Extraordinaire ("GLIMPSE"), wherein 220 square degrees the galactic plane was surveyed at a pixel size of 0.6 arcsec, and to a pointing accuracy of ~ 0.3 arcsec. The observations were undertaken using the Infrared Array Camera (IRAC; Fazio et al. 2004), and employed filters having isophotal wavelengths (and bandwidths $\Delta\lambda$) of 3.550 $\mu$m ($\Delta\lambda$ = 0.75 $\mu$m), 4.493 $\mu$m ($\Delta\lambda$ = 1.9015 $\mu$m), 5.731 $\mu$m ($\Delta\lambda$ = 1.425 $\mu$m) and 7.872 $\mu$m ($\Delta\lambda$ = 2.905 $\mu$m). The normal spatial resolution for this instrument varies between ~1.7 and ~2 arcsec (Fazio et al. 2004), and is reasonably similar in all of the bands, although there is a stronger diffraction halo at 8 $\mu$m than in the other IRAC bands. This leads to differences between the point source functions (PSFs) at ~0.1 peak flux.

We have used these data to produce a colour coded image of Sh 2-89, where 3.6 $\mu$m is represented as blue, 4.5 $\mu$m as green, and 8.0 $\mu$m is indicated by red (see Fig. 1). This image was also processed using unsharp masking techniques, whereby a blurred or "unsharp" positive of the original image is combined with the negative. This leads to a significant enhancement of higher spatial frequency components, and an apparent "sharpening" of the image (see e.g. Levi 1974).

Finally, contour maps have been produced in the differing IRAC bands, where the emission $E_n$ for contour level *n* is given through $E_n = A10^{(n-1)B}$ MJy sr$^{-1}$, and contour parameters *[A,B]* are cited in the caption to Fig. 2. Similarly, profiles through the source are given in Fig. 3, a process which required correction for the effects of background emission in the differing photometric bands. These components are particularly strong at 5.8 and 8.0 $\mu$m, where they also result in slight gradients in background of order of 5 10$^{-4}$ MJy/sr/pix (although these gradients also depend upon the direction of the slice). We have removed these trends by subtracting lineal ramps from the results – a procedure which is more than adequate given the limited size of the source.

## 3. The Environment of Sh 2-89

### 3.1 Mid-Infrared Imaging

A colour image of Sh 2-89 is illustrated in Fig. 1, as described in Sect. 2. It is clear from this that the highest surface-brightness portions of the source extend over a range of ~ 1.5 arcmin, and that the brighter nuclear region has a size of ~15 arcsec; values which correspond to dimensions of ~ 1.3 and 0.2 pc at the distance D $\cong$ 3.0 kpc cited above (Sect. 1). Much of the action in this source is centred close to $\alpha$(2000.0)

= 19h 50m 00s, $\delta$(2000.0) = 26° 27' 37", however, where there is evidence for the intersection of two parabolic emission arcs (see also the mapping to be described below). A variety of observations have identified sources in this regime, certain of which are indicated in Fig. 1. They include the MIR detection of a compact cluster (Mercer et al. 2005; indicated using a large white circle, the diameter of which is equal to that quoted for the cluster); a far infrared (FIR) source IRAS 19479+2620 (Illustrated using the yellow error ellipse); and an X-ray source AX J1950.1+2628 (Sugizaki et al. 2001). Apart from these, various independent radio observations have resulted in differing estimates for the centroid of ionised emission, and these are indicated using the three blue crosses (the coordinates for RFS 1028 are taken from Reich et al. 1984; of MITG J195001+2627 from the Merged MIT-Green Bank 5GHz Survey of Bennet and collaborators, available as a VizieR catalogue at http://vizier.u-strasbg.fr/viz-bin/VizieR; and of Sh 2-89 are taken from Taylor et al. 1996). An investigation of the MIR colour indices of sources within this field, combined with the colour-colour YSO modeling of Allen et al. (2004), also enables us to identify three likely examples of Class I YSOs. These are identified using green circles in Fig. 1, whence it will be noted that one of these (the brightest) is located close to the nucleus of the source. The colour indices of this object are [3.6]-[4.5] = 0.707 mag, and [5.8]-[8.0] = 0.535 mag, whilst fluxes and positions for all of the stars are summarised in Table 1. These three sources are identifiable because photometry is available in all four of the IRAC bands; it is likely however that many other Type I sources exist, most of them having only partial photometric results.

Two sources are observed close to the centre of Sh 2-89. The first (which we shall indicate as Star B) has colour indices [3.6]-[4.5] = 0.629 mag, [5.8]-[8.0] = 0.905 mag characteristic of Class II YSOs – although an earlier (Class I) phase of evolution cannot entirely be excluded; the source lies close to the Class I/Class II interface defined by Allen et al. (2004). A further source (star A) is also located close to the centre of Sh 2-89, and represents the putative symbiotic star identified by Corradi et al. (2008). It will be further described in Sects. 3.4→3.6 below.

Finally, we are able to identify a further seven Class II YSOs within the central 7.5x7.5 arcmin$^2$ of the region. Details for these sources (as well as the photometry for stars A are B) are to be found in Table 1. The source IRAS 19479+2620 may also represent a YSO, although since significant fluxes are only available in the 25 μm band, and there is no obvious counterpart at longer and shorter wavelengths, its nature remains unclear. Similarly, images at 8.28 μm, 12.13 μm, 14.65 μm, and 21.34 μm, acquired during the Spitzer MIPSGAL project (Carey et al. 2009), appear to show both the bright central nucleus and (at shorter wavelengths) a variety of stars within the field – although none of these appear to correspond to the YSOs cited above. It is possible however that the central star B, and the adjacent Class I YSO have fluxes which are merged with, and indistinguishable from that of the central nebular structure.

### 3.2 MIR Mapping and Spectroscopy of Sh 2-89

In addition to the combined colour map illustrated in Fig. 1, we have also produced the separate MIR contour maps illustrated Fig . 2. These show an evolution of the structure from 3.6 μm, where a single arc is visible, and the morphology is similar to that observed in Digital Sky Survey images; through to 8.0 μm, where the emission is seen to extend over the entirety of the field. Note, in particular, the presence of two arc-like structures at 8.0 μm, close to the bright centre of the source, also indicated using the white curves in the insert in Fig. 3. These are likely to represent intersecting ionisation fronts (see below), and suggest a bipolar configuration.

Although some of this evolution in morphology is related to the high levels of reddening in the source (see e.g. our further discussion in Sect. 3.3 below), and enhanced extinction at 3.6 and 4.5 μm, it is also likely to represent a real evolution in the properties of the region, as differing sources of emission come into play at longer MIR wavelengths.

We have undertaken visual spectroscopy along a direction which passes through star A, and also includes several other important elements of the nebular structure. The location of the slit is indicated in Fig. 3, where we have superimposed profiles for H$\alpha$, [NII] $\lambda$ 6583 Å, [SII] $\lambda\lambda$ 6716+6731 Å, and the four IRAC bands. Several interesting features immediately stand out.

In the first place, it is clear that the intensity of MIR emission increases with increasing IRAC wavelength, with ratios 8.0µm/4.5µm, 5.8µm/4.5µm and 3.6µm/4.5µm taking respective mean values of ≈25, 10 and 1.3 These ratios are typical of what is observed in other compact HII regions (Phillips & Ramos-Larios 2008), and bipolar structures associated with YSOs (Phillips & Perez-Grana 2009), and appear reasonably invariant over the projected area of the source. They are likely, as in these other sources, to reflect the importance of warm dust continua, and (above all) strong emission due to polycyclic aromatic hydrocarbons (PAHs). Although the 4.5 µm band is not known to contain PAH type features – it is normally dominated by ionic and molecular transitions, and the Bremsstrahlung continuum – such PAH bands and their associated plateaus are important at 3.3, 6.2, 7.7, and 8.6 µm (see e.g. Tielens 2005). This leads to enhanced emission in the 3.6, 5.8 and 8.0 µm photometric bands – although the importance of 3.3 µm PAH emission is normally quite modest.

The H$\alpha$, [NII] and [SII] profiles, by contrast, show markedly differing structures. It will be noted for instance how the H$\alpha$ emission peaks close to relative position (RP) $\cong$ 18.1 arcsec, and declines strongly to larger values of RP (where RP = 0 arcsec corresponds to the position of Star A). By contrast, the 8.0 µm flux increases strongly at the position of the H$\alpha$ peak, and thereafter forms a broad ridge of emission extending up to ~ 40 arcsec (and peaking close to $\cong$ 31 arcsec). Similar tendencies are noted at 5.8 µm as well. All of these trends are comprehensible where we are observing the edge-on projection of an ionisation front, wherein the peak in H$\alpha$ emission lies close to an HI/HII interface. The neutral region defined by the PAH emission ridge may take the form of molecular gas, or represent a photo-dissociative regime (PDR) within which $H_2$ is dissociated by FUV photons (6 ev $\leq$ h$\nu$ $\leq$ 13.6 ev).

It is of interest, in this respect, to note that the lower excitation [NII] and [SII] lines have similar profiles to that of H$\alpha$ – but are much less strongly peaked, and appear to extend further into the MIR/PAH emission ridge. This is broadly as might be anticipated where there is ionisation stratification.

It is important, in such a discussion, to emphasise that we are considering the two-dimensional projection of a complex three-dimensional structure, and this will inevitably limit our ability to comprehend what is happening in this regime. Nevertheless, there seems little doubt that our analysis of this region, where the slit crosses one of the broad emission arcs, is indicating that the butterfly pattern of visual emission is associated with local I-front activity, where the UV radiation field and/or winds are impacting neutral arm-like structures. It is also clear that I-fronts are not limited to this regime alone, as is testified by the similar changes in profile noted at RP = -49 arcsec, and the narrow emission rims evident in Fig. 1.

A particular fascinating aspect of these traverses is the sharp peaking of emission at RP = 0 arcsec. This corresponds to the position of star A in Fig. 1, and the putative symbiotic source of Corradi et al (2008). This source appears to be strong in all of the MIR bands, and is also dominant in H$\alpha$. The slight off-centre peak in transitions of [SII] is an artefact of the observations (see our further discussion in the following section), and is not indicative of enhanced stellar emission. The spectrum of this star will be the subject of further analysis in Sects. 3.4→3.6.

### 3.3 Densities and Extinction in Sh 2-89

There are at least two physical characteristics which may be deduced from the results described above. In the first place, it seems that the [SII] $\lambda6716/\lambda6731$ ratios are broadly consistent along the slit, and over the range of distances ~ 30 arcsec over which this parameter may be reliably determined. We determine a value $<\lambda6716/\lambda6731> = 1.16 \pm 0.42$, where the quoted error represents the standard deviation in the results (i.e. the measured scatter in the estimates of this ratio). For an assumed mean temperature close to ~$10^4$ K, and adopting the [SII]/density calibration of Stanghellini & Kaler (1989), this then implies an overall density $n_e \cong (3.7 \pm 1.0) \, 10^2$ cm$^{-3}$.

The estimation of extinction is particularly problematic, since levels of H$\beta$ emission are low. For the combined nebular spectrum to the NW of star A, for instance, we determine an H$\beta$ flux F(H$\beta$) = (5.78 ± 1.88) $10^{-16}$ ergs cm$^{-2}$ s$^{-1}$, for instance, for which case H$\alpha$/H$\beta$ ratios imply extinctions $A_V = 6.2 \pm^{1.0}_{0.7}$ mag; where have assumed that intrinsic H$\alpha$/H$\beta$ ratios correspond to case B conditions, taken electron temperatures $T_e = 10^4$ K, and adopted densities $n_e \approx 10^2$ cm$^{-3}$ based on the analysis presented above. We have also adopted the standard extinction curve described by Osterbrock (1989).

Whilst this value of extinction is therefore pretty uncertain, it does appear more-or-less consistent with the values of $E_{B-V} \cong 1.62$ determined by Forbes (1989) – an estimate which would imply $A_V \cong 4.9$ where one adopts the ratio $R = A_V/E_{B-V} = 3.0$ of Forbes.

For the star itself, on the other hand, we obtain $F(H\beta) = (2.69 \pm 1.30)$ $10^{-16}$ ergs cm$^{-2}$ s$^{-1}$ – a result which is at the limits of being significant. The resulting extinctions would be of order $A_V \sim 8.5 \pm^{1.7}_{1.1}$; a value which is larger than estimated by Forbes (1989) for other stars in the region, and perhaps indicative significant levels of local stellar reddening.

### 3.4 The Nature of Star A

We have noted, in Sect. 2, that in addition to the broad components of nebular emission described in Sect. 3.1 and 3.2, we have also determined the spectrum of the symbiotic candidate of Corradi et al. (2008) (star A in Fig. 1), which is illustrated in Fig. 4.

It is apparent, from this, that the star reveals evidence for dips in its spectrum which are likely indicative of TiO absorption bands. There is however no clear evidence for [HeII] $\lambda$ 4686 Å or [OIII] $\lambda$ 5007 Å emission, such as would be expected where the source was symbiotic, as suggested by Corradi et al. (2008).

A further interesting feature is the apparent presence of H$\alpha$, [NII] and [SII] emission, although some care must be taken in the interpretation of these transitions. Whilst we have been careful in removing background elements of emission – emission associated with the enveloping HII region Sh 2-89 (see Sect. 2) – it is possible that the weak spike corresponding to [NII] $\lambda$ 6583 Å represents contamination of the stellar flux. Similarly, we note that the presence of "bad" pixels close to the position of the [SII] doublet tend to distort and amplify the apparent flux of these transitions; they are also, for the most part, impossible the remove from the present results. The best that one can therefore state is that [NII]/H$\alpha$ ratios are of order $\leq$ 4.7, and [SII]/H$\alpha$ ratios are $\leq$ 5.3.

The most striking feature of the spectrum, however, is the very strong level of H$\alpha$ emission, implying overall flux levels of order 1.01 10$^{-13}$ ergs cm$^{-1}$ s$^{-1}$. It seems clear that the star is either emitting a strong stellar wind, or is associated with an extremely compact HII region – a region which perhaps derives from the ionisation of a circumstellar disk or envelope. Before addressing this question further, however, it is as well to establish what the nature of the star might be.

It would seem, from our present spectroscopy, that the identification of Corradi et al. may be in error. However, it is difficult to establish that this is definitively the case with the evidence to hand. To see why this is so, note that the 2$\sigma$ upper limit ratios I(HeII)/I(H$\alpha$) and I([OIII])/I(H$\alpha$) are of respective order ~ 0.044 and ~ 0.046. It is instructive to compare these ratios with values obtained for other (reddened) symbiotic spectra.

Four of the five D-type symbiotics investigated by Pereira, Landaberry & Junqueira (1998) have I(HeII)/I(H$\alpha$) and I([OIII])/I(H$\alpha$) ratios which are significantly greater than these limits, for instance – although the estimates for He 2-38 are comparable, and of respective order 0.055 and 0.042. The broader symbiotic selection of Mikolajewska, Acker & Stenholm (1997), on the other hand – a sample which is dominated by S-type sources (~80% of the sample) – indicates that fully 60% have [OIII]/H$\alpha$ line ratios which are comparable to or less than the limits cited above.

It follows that given the limits imposed by the S/N of our present spectral data, and the likely levels of ISM extinction determined above, it is difficult to exclude the possibility that the star is symbiotic on the basis of the visual data alone. Similar conclusions also apply for the infrared data present above. Whilst S-type symbiotics tend to have relatively weak MIR-FIR emission levels, leading to lower fluxes than those observed in the NIR 2.2 $\mu$m band, for instance (see e.g. Kenyon, Fernandez-Castro & Stencel 1988), D-type sources have much more strongly emitting cool enveloping shells, and spectral distributions which are sometimes comparable to that noted here (see below, and Kenyon, Fernandez-Castro & Stencel 1986, 1988).

It follows that neither the D- nor S-type symbiotics are entirely capable of explaining the infrared and visual trends – D-types appear, for the most part, to have higher [OIII] and HeII line intensities, whilst S-types have much weaker IR emission. However, the D-type option cannot be entirely ruled out on the basis of the information presented above.

On the other hand, the presence of an extended, highly complex, and filamentary HII region, with indications for associated star formation, CO emission, and compact clusters of stars (see Sect. 3.1) rather suggests that star A is better classified as a possible YSO; a suggestion which is further explored in our modelling of this source in Sect. 3.5 below.

## 3.5 Infrared and Visual Photometry and Modelling of Star A

Spitzer and 2MASS photometry, and photometry based upon our present spectroscopy (from which we have determined B & V band fluxes) is presented for star A in Fig. 5 (top panel), where the results are fitted with a series of best fit YSO models. The infrared results are summarised in Table 1, whilst the V band magnitude is found to be V = 17.00 ± 0.05, and the B band estimate is 18.72 ± 0.07.

We have, for the purpose of the modelling, made use of the ~2 $10^5$ synthetic YSO SEDs evaluated by Robitaille et al. (2006), analysed using the SED fitting tool developed by Robitaille et al. (2007), in which continua have been determined for differing dust and gas geometries, varying dust properties, and using 2D radiative transfer modelling for a large region of parameter space. The fitting of the models also takes account of the apertures employed for the photometry. Given that our present source does not appear to be resolved in any of the wavebands, we have set this parameter to be equal to the FWHM of the PSFs in each of the respective channels – ranging from ~2 arcsec for the IRAC and visual wavebands, to ~5 arcsec for the 2MASS results.

The best fit model corresponds to a $\chi^2$ of 30.08 (referred to below as $\chi_{BEST}^2$), and is indicated by the solid black curve. We also show other models for which $\chi^2 - \chi_{BEST}^2 < 3$ (grey curves). We have finally specified, for these models, that stellar distances must be in the range $2 < D/kpc < 4$, and IS extinctions of order $4 < A_V < 7$ (see e.g. Sects. 1 and 3.3).

In addition to these trends, we have also indicated the differing components of flux which make up the model YSO spectrum (lower panel of Fig. 5). In this case, the total flux is indicated as black, the stellar flux is blue, the stellar photospheric flux is indicated by the dashed line (this is the flux prior to reddening by circumstellar dust); the disk flux is green; the scattered flux is yellow; the envelope flux is red; and the thermal flux is orange. Unless otherwise stated, the results include the effects of circumstellar extinction, but not of IS extinction. They also assume a representative distance of 1 kpc, and an aperture of close to 5 arcsec.

It is apparent, from this latter modelling, that the continuum can be represented by a star in which most of the MIR/NIR emission is derived from the central star and disk components of emission. Envelope fluxes are of most importance at $\lambda > 10$ μm, although it appears that they are likely to be weak. It therefore seems that star A can best be modelled in terms of a Class II YSO.

Given that this interpretation is correct, it is then of interest to determine what this might imply for the physical properties of the source. Our results are illustrated in Fig. 6, and again apply for models in which $\chi^2 - \chi_{BEST}^2 < 3$. The grey portions of these graphs indicate the relative numbers of models occurring in the ranges of parameter space investigated here.

It is apparent, from these results, that the properties of the star are reasonably well defined. The stellar age $T_{EV}$ is likely to be in the range $2 \cdot 10^6 \rightarrow 10^7$ years; stellar extinction $A_V(STAR)$ (i.e. the extinction caused by dust about the star) is <1 mag; luminosities $L_*$ are required to be $1 \rightarrow 3 \cdot 10^2$ $L_\odot$; the stellar mass $M_*$ is $3 \rightarrow 5$ $M_\odot$; and the present photospheric radius of the star $R_*$ is of the order of $\sim 2 \rightarrow 3$ $R_\odot$. Finally, the stellar temperature $T_{EFF}$ is $\sim 10^4 \rightarrow 2 \cdot 10^4$ K.

So it seems likely that we are dealing with an intermediate mass star with a reasonably high luminosity, but a star which still has too low a temperature to appreciably ionise its surroundings. The TiO absorption bands would presumably arise in cool and neutral gas, located well above the photospheric region – or imply that the effective temperature of the star is even lower than determined from our modelling above.

It is finally worth remarking that the top ten YSO models imply a distance $2.00 < D/kpc < 2.29$, and interstellar extinction $5.12 < A_V < 5.31$; values which are, given the uncertainties, respectably close to those determined by Forbes (1989) and other authors (see Sects. 1 & 3.3).

Where one loosens the constraints on the models, and permits (say) that $0 < A_V < 50$, and $0.012 < D/kpc < 10$, then it turns out that the range of YSO parameters are very similar to those quoted above (specifically $10^6 < T_{EV}/yrs < 10^7$; $2 < M_*/M_\odot < 4$; $2 < R_*/R_\odot < 3$; $50 < L_*/L_\odot < 300$; $10^4 < T_{EFF}/K < 2\,10^4$). The IS extinction, for which we determine values $A_V \sim 4.85 \rightarrow 5.44$ mag for the top ten models, is also surprisingly similar to those determined from observations (see e.g. Sect. 3.3). Although most model stellar extinctions are again very low, and in the range cited above, a few models also have values $A_V(STAR)$ extending as high as $10 \rightarrow 500$ mag – values which may be more consistent with the high H$\alpha$/H$\beta$ ratios observed within the central star (see Sect. 3.3). Loosening the constraints therefore leads to predicted extinctions and distances which are consistent with those deduced from the observations cited above.

### 3.6 The H$\alpha$ Emission in Star A

Given, from Sect. 3.5, that star A is probably a Class II YSO, it is of interest to return to the interpretation of what might be responsible for H$\alpha$ emission in this source.

HI line emission in YSOs is thought to arise from a variety of causes, including chromospheric activity, jets, winds deriving from the circumstellar disks, Herbig-Haro activity and the like. The outflows in moderate to high mass YSOs, such as appears to be the case for Star A (see Sect. 3.5), are often thought to be a product of the collapse process itself, and to arise as a result of radiation pressure and magnetic fields (Yorke & Sonnhalter 2002; Banerjee & Pudritz 2007). For such cases, the infrared HI emission from radiation driven accretion disk winds has been analysed in considerable detail by Sim et al. (2005). To gain an idea of the order-of-magnitude mass-losses which are implied by the present Hα fluxes, we initially shall assume that the Hα emission is optically thin, spherically symmetric, is emitted under case B conditions (see e.g. Osterbrock 1989), and that temperatures $T_e$ are of order $10^4$ K. Where densities fall-off radially as

$$n_e = n_0 \left(\frac{R_0}{R}\right)^\beta \quad \ldots\ldots(1)$$

where $R_0$ is the internal radius of the wind, then the total Hα flux would be given through

$$I(H\alpha) \cong 2\pi\alpha_H n_0^2 R_0^3 \left[\frac{2}{(2\beta-3)} - \left(\frac{R_*}{R_0}\right)^2 \frac{1}{(2\beta-1)}\right] \quad \ldots\ldots(2)$$

where both here and in what follows we allow for occultation of components of the flux by the central star, and $\alpha_H$ is the volume emission coefficient for Hα. This relation is valid providing that $\beta > 1.5$. Where $R_0 = R_*$ then this gives an observed flux of

$$I_{OBS}(H\alpha) \cong \frac{\alpha_H n_0^2 R_*^3}{2D^2} \frac{(2\beta+1)}{(2\beta-3)(2\beta-1)} \quad \ldots\ldots(3)$$

an expression which can be shown to imply corresponding mass-loss rates

$$\frac{dM}{dt} \cong 6.46\,10^{-2}\,\mu\xi^{-1}\left(\frac{R_*}{R_{SUN}}\right)^{0.5}\left(\frac{V_w}{km\,s^{-1}}\right)\left(\frac{D}{kpc}\right)\left[\left(\frac{I_{OBS}(H\alpha)}{erg\,cm^{-2}\,s^{-1}}\right)\frac{(2\beta-3)(2\beta-1)}{(2\beta+1)}\right]^{0.5} M_{SUN}\,yr^{-1} \quad \ldots\ldots(4)$$

where $V_W$ is the velocity of the wind (assumed to be constant), D is the source distance, $\mu$ is the mean atomic mass of the gas per H atom (taken to be 1.42), and $\xi$ is the mean number of electrons per H atom (~1.21).

Substituting values $\beta = 2$, $V_W = 10^2$-$10^3$ km s$^{-1}$, $R_* = 10$ $R_\odot$ (see Sect. 3.5), $D = 3.0$ kpc (sect. 1), and the extinction corrected values of $I_{OBS}(H\alpha)$ cited above, one then obtains mass-loss rates ~ 4.5 10$^{-8}$ - 4.5 10$^{-7}$ $M_\odot$ yr$^{-1}$ which appear reasonable for a star at this stage of its evolution. Similar values would also be deduced where levels of flow collimation are higher (see e.g. Beuther et al. 2002; Davis et al. 2004), although densities would be required to be significantly larger.

## 4. The Environment of Sh 2-89

It is clear, from the discussion above, that the environment of Sh 2-89 is relatively complex. There is evidence for arcing filaments centred close to star B in Fig. 1, where the morphology appears similar to that in many bipolar outflows. Profiles in the MIR and visible show that the visible arcs correspond to ionisation fronts, in which peaks in the projected H$\alpha$ emission occur at an interface with neutral (molecular or photo-dissociated) gas. There is also evidence that lower excitation [NII] and [SII] transitions may penetrate deeper into the PAH emitting regimes – an indication of likely ionisation stratification at the HI/HII interface.

Star formation appears to have been occurring over an extended period of time – and is likely still to be taking place. Thus Mercer et al. (2005) have noted the presence of a compact MIR cluster close to the centre of the source, within which there may even be evidence for an X-ray emitting source (Sugizaki et al. 2001). The *SST* MIR photometry, combined with the colour-colour analysis of Allen et al. (2004), suggests the presence of three Class I YSOs - one of which, the brightest, is located close to highest surface brightness regions of the source. It is likely that many more Class I sources litter this regime, but have yet to be detected because they are either faint, or MIR photometry is incomplete.

There is, apart from this, evidence for several Class II YSOs – a census of the MIR colours of point sources in this regime suggest that there are at least nine such objects within the map area illustrated in Fig. 1 (see Table 1, and Sect. 3.1). Two of these (A & B) are identified in Fig 1 – with source A representing an object of particular interest for this study.

Our spectroscopy of the star shows it to possess clear TiO absorption bands – although there is, apart from this, little evidence to suggest that it may be a symbiotic source. An analysis and modelling of visual, 2MASS and *SST* photometry suggests rather that we may be dealing with a Class II YSO. The luminosity is likely to be appreciable (~$0.5 \rightarrow 3 \ 10^2 \ L_\odot$), although current photospheric temperatures are modest (~$10^4 \rightarrow 2 \ 10^4$ K).

We have noted that the large H$\alpha$ intensities associated with the star are consistent with a moderate mass-loss rate stellar/disk wind or jets; components which may arise due to X-type flows (e.g. Shang et al. 1998; Arce & Sargent 2006; Shu et al. 1988, 1994, 2000; Ostriker & Shu 1995), D winds (Matt et al. 2003; Pudritz et al. 2007) and the like.

The presence of broad ionisation fronts within Sh 2-89 is also consistent with moderately high energy densities of UV ionising photons. Given a typical diameter for Sh 2-89 of 1.3 pc (Sect. 3.1), for instance, and electron densities in the region of ~ $3.7 \ 10^2$ cm$^{-3}$ (Sect. 3.3), then this would imply the need for ~$1.3 \ 10^{48}$ H ionising photons/s – a flux that may arise from a single O8.5V type star, an early Type III B supergiant, or multiple lower temperature/luminosity stars (see e.g. Martins et al. 2005). This extended radiation field may be interacting with neutral structures about star A, and causing limited levels of ionisation.

Finally, we note that the morphology of this region suggests the presence of a nuclear bipolar configuration, perhaps arising from stellar winds deriving from the local YSO population. This more generalised wind may presumably be interacting with the disk and/or envelope of the star as well.

It is therefore apparent that a variety of external agents may be important in ionising material close to star A, and leading to the H$\alpha$ emission noted in our spectrum in Fig. 4.

## 5. Conclusions

We have undertaken a detailed analysis of the HII region Sh 2-89, investigating the properties of the source in the MIR, and undertaking spectroscopy of the interior portions of the nebula. We have noted that the large scale structure of the source includes several extended emission arcs, various of which appear to intersect in the bright nucleus of the source. The morphology in this regime is reminiscent of those for certain young bipolar flows, such as have previously been discussed by Phillips & Perez-Grana (2009). This core is, in turn, associated with a variety of stellar components, the brightest of which appear to be identified as Class I and II YSOs.

One of these stars has been previously identified as a candidate symbioic star. However, although our spectroscopy indicates the presence of TiO bands, there are no corresponding higher excitation or forbidden lines. It is pointed out that the apparent absence these lines cannot be used to definitively discriminate against a symbiotic identification. However, the spatial context of the star, whereby it appears to be located within a compact HII region with much evidence for recent star formation; and the results of YSO modelling of the visual, MIR and NIR continuum of this object, suggests that it may be a Class II YSO with temperature $T_{EFF}$ ~$10^4$-$210^4$ K and luminosity ~ 0.5→3 $10^2$ $L_\odot$. Most of the MIR emission appears to be associated with the accretion disk.

This star is also associated with strong H$\alpha$ emission, the nature of which is not entirely clear. It is possible for instance that it is associated with radiatively accelerated stellar winds, in which case mass-loss rates would be of order ~4.5 $10^{-7}$ $M_\odot$ yr$^{-1}$ (for wind velocities $V_W$ of $10^3$ km s$^{-1}$; the value would be proportionately smaller where $V_W$ is less). It is also possible that much of the emission may be associated with local structures, such as the circumstellar disk and/or surrounding YSO envelope. Ionisation of these components may occur through collisional ionisation by stellar or nebular winds, and/or through the influence of the pervading nebular UV ionisation field.

Apart from this, the census of YSOs, as determined from MIR and FIR photometry, suggests that there may be at least three Class I YSOs, and multiple Class II YSOs – although our total for these sources is likely to represent a severe lower limit. There is also evidence for a MIR cluster of stars detected by Mercer et al. (2005). It is therefore clear that star formation is particularly active in this regime, and likely to include a broad range of stellar evolutionary states.

Finally, we have obtained profiles through the extended arcs in Sh 2-89. Our MIR results show that much of the emission is likely to be associated with PAH bands and plateau features, suggested that much of the material is molecular, or concentrated in photo-dissociative regimes. It is noted that the H$\alpha$ profiles tend to peak to one side of the arcs, as would be expected where we are detecting ionisation fronts at an HI/HII interface. Similarly, there is evidence for stronger penetration of the neutral regimes by lower excitation transitions – a feature which might be expected where there is appreciable and spatially resolved ionisation stratification.

## Acknowledgements

We thank an anonymous referee for several useful suggestions which greatly improved the analysis in this paper. This research is based, in part, on observations made with the Spitzer Space Telescope, which is operated by the Jet Propulsion Laboratory, California Institute of Technology under a contract with NASA. Support for this work was provided by an award issued by JPL/Caltech. In addition to this, the work makes use of data products from the Two Micron All Sky Survey, which is a joint project of the University of Massachusetts and the Infrared Processing and Analysis Center/California Institute of Technology, funded by the National Aeronautics and Space Administration and the National Science Foundation. The 2MASS data was acquired using the NASA/ IPAC Infrared Science Archive, which is operated by the Jet Propulsion Laboratory, California Institute of Technology, under contract with the National Aeronautics and Space Administration. GRL acknowledges support from CONACyT (Mexico) grant 93172.

TABLE 1

2MASS and Spitzer Photometry for Class I & II YSOs within Sh 2-89

| SPITZER I.D. SSTGLMC | RA (J2000) H:M:S | DEC (J2000) D:M:S | CLASS[3] | J mag | σ(J) mag | H mag | σ(H) mag | $K_S$ mag | σ($K_S$) mag | [3.6] mag | σ([3.6]) mag | [4.5] mag | σ([4.5]) mag | [5.8] mag | σ([5.8]) mag | [8.0] mag | σ([8.0]) mag |
|---|---|---|---|---|---|---|---|---|---|---|---|---|---|---|---|---|---|
| G062.9616+00.1791 | 19:49:48.77s | 26:32:22.87s | II | 15.60 | 0.08 | 13.98 | 0.05 | 13.39 | 0.05 | 12.97 | 0.07 | 12.85 | 0.10 | 12.93 | 0.33 | 12.30 | 0.23 |
| G062.9176+00.0981 | 19:50:01.49s | 26:27:38.14s | II | 13.13 | 0.02 | 11.74 | 0.02 | 10.62 | 0.02 | 9.33 | 0.04 | 8.79 | 0.05 | 8.23 | 0.04 | 7.44 | 0.04 |
| G062.9372+00.1559 | 19:49:50.79 | 26:30:24.79 | I | … | … | … | … | … | … | 15.35 | 0.18 | 14.06 | 0.16 | 12.81 | 0.28 | 12.31 | 0.17 |
| G062.9126+00.1056 | 19:49:59.08 | 26:27:36.43 | I | … | … | … | … | … | … | 8.38 | 0.04 | 7.67 | 0.05 | 7.08 | 0.04 | 6.54 | 0.04 |
| G062.8753+00.0829 | 19:49:59.22s | 26:24:59.34s | II | … | … | … | … | … | … | 13.49 | 0.07 | 13.13 | 0.14 | 12.45 | 0.21 | 11.63 | 0.07 |
| G062.9133+00.1031[1] | 19:49:59.74 | 26:27:34.03 | II | … | … | … | … | … | … | 8.32 | 0.05 | 7.69 | 0.05 | 6.97 | 0.05 | 6.07 | 0.15 |
| G062.9176+00.0981[2] | 19:50:01.49 | 26:27:38.14 | II | 13.13 | 0.02 | 11.74 | 0.02 | 10.62 | 0.02 | 9.33 | 0.04 | 8.79 | 0.05 | 8.23 | 0.04 | 7.44 | 0.04 |
| G062.9383+00.0809 | 19:50:08.28s | 26:28:10.80s | II | 13.65 | 0.026 | 12.79 | 0.029 | 12.00 | 0.03 | 10.72 | 0.06 | 10.15 | 0.04 | 9.66 | 0.05 | 9.11 | 0.04 |
| G062.9980+00.1089 | 19:50:09.94s | 26:32:07.22s | II | 15.14 | 0.05 | 13.76 | 0.04 | 13.21 | 0.04 | 12.68 | 0.057 | 12.52 | 0.092 | 12.46 | 0.22 | 12.06 | 0.18 |
| G062.9793+00.0863 | 19:50:12.63 | 26:30:27.78 | I | 16.53 | 0.15 | 15.25 | 0.10 | 14.37 | 0.10 | 13.45 | 0.08 | 12.74 | 0.11 | 12.26 | 0.16 | 10.76 | 0.11 |
| G062.9078+00.0402 | 19:50:13.52s | 26:25:21.85s | II | 14.88 | 0.05 | 13.37 | 0.03 | 12.64 | 0.03 | 12.19 | 0.06 | 12.00 | 0.09 | 12.98 | 0.12 | 11.67 | 0.18 |
| G062.9801+00.0802 | 19:50:14.13s | 26:30:19.14s | II | 15.47 | 0.07 | 13.20 | 0.03 | 12.23 | 0.03 | 11.57 | 0.05 | 11.49 | 0.08 | 11.19 | 0.09 | 10.77 | 0.13 |

Footnotes: [1]Star B; [2]Star A; [3]Putative Class estimate for the YSOs

# Figure Captions

**Figure 1**

Imaging of Sh 2-89 with the *SST*, where we combine results at 3.6 $\mu$m (represented as blue), 4.5 $\mu$m (green), and 8.0 $\mu$m (red). We have also superimposed the positions of the MIR cluster detected by Mercer et al. (2005) (white circle – the size of which corresponds to the quoted dimensions of the cluster); the X-ray source AXJ1950.1+2628 (Sugizaki et al. 2001); the FIR point source IRAS 19479+2620 (indicated by the yellow error ellipse); and a variety of radio centroids (blue crosses) corresponding to observations taken by Taylor et al. (1996) (Sh 2-89), Bennet and his collaborators (MITG J195001+2627; see Sect. 3.1 for further details) and Reich et al. (1994) (RFS 1028). We have, apart from these, also indicated the positions of stars A & B referred to in the text, and of three Class I YSOs (green circles).

**Figure 2**

Contour maps of Sh 2-89 in the four IRAC photometric channels, where it will be seen that the intensity and extension of the source increases with increasing MIR wavelength. The contours are set at logarithmic intervals, and have contour parameters (A,B) given by (1.9, 0.1628) at 3.6 $\mu$m; (1.4, 0.1763) at 4.5 $\mu$m; (7.9, 0.1146) at 5.8 $\mu$m; and (20.5, 0.0765) at 8.0 $\mu$m.

**Figure 3**

MIR and visual profiles through the centre of Sh 2-89, where we illustrate trends for H$\alpha$, [SII], [NII] and the four IRAC channels. The orientation of the traverse is indicated in the inserted figure, and corresponds to a width of 2.3 arcsec and length of 295 arcsec, whilst the vertical scale refers to the MIR results. The calibration of the visual results is not provided here, although the profiles have the correct relative intensities. Note the presence of star A at the centre (RP = 0 arcsec), responsible for intense peaks in the visual and MIR profiles. We have also indicated the positions of field stars 1 and 2 in the inserted image, where the primary I-fronts are delineated using white contours. These I-fronts are also evident in the profiles at RPs of +18 and -49 arcsec.

**Figure 4**

Spectrum of star A, where we have also indicated certain of the more interesting absorption/emission components.

**Figure 5**

YSO modelling of our present visual, *SST* and 2MASS photometry for star A. The upper panel shows photometric fitting for models having $\chi^2 - \chi^2_{BEST} < 3$; where the best-fit solution is indicated by the black curve. The dashed curve represents the stellar flux reddened by IS extinction, but not taking account of circumstellar extinction. The lower panel shows the various emission components making up this model, details of which are indicated in the text. Suffice to remark that the stellar component (blue curve) dominates at wavelengths $< 1$ μm, and the circumstellar disk (green curve) is important for wavelengths between 1 and 10 μm.

**Figure 6**

The ranges of physical parameter determined from YSO modelling of star A, where the grey bars indicate the relative densities of model solutions searched in this analysis, and the cross-hatched bars correspond to the values determined for our best fit solutions. All of the graphs are normalised to unity.

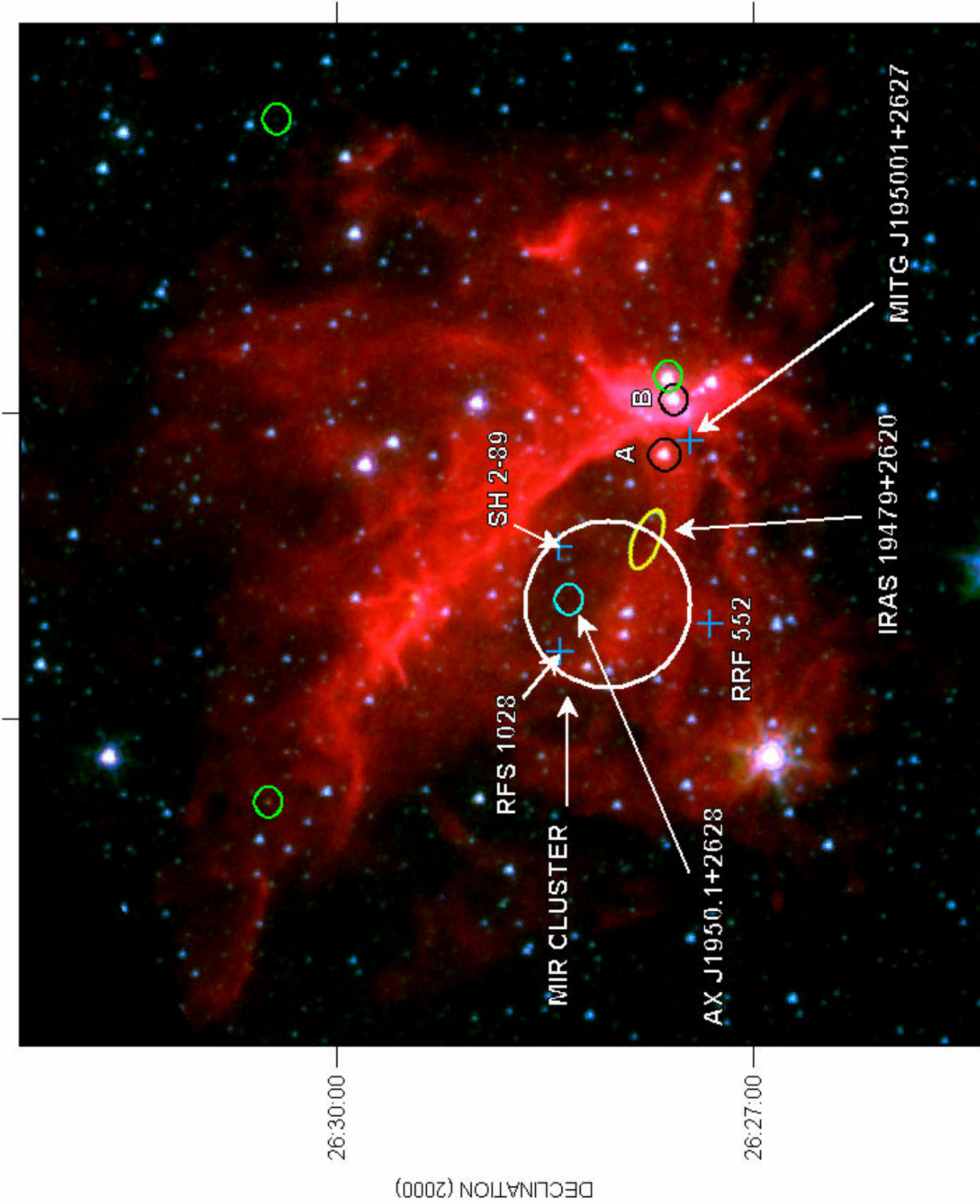

FIGURE 1

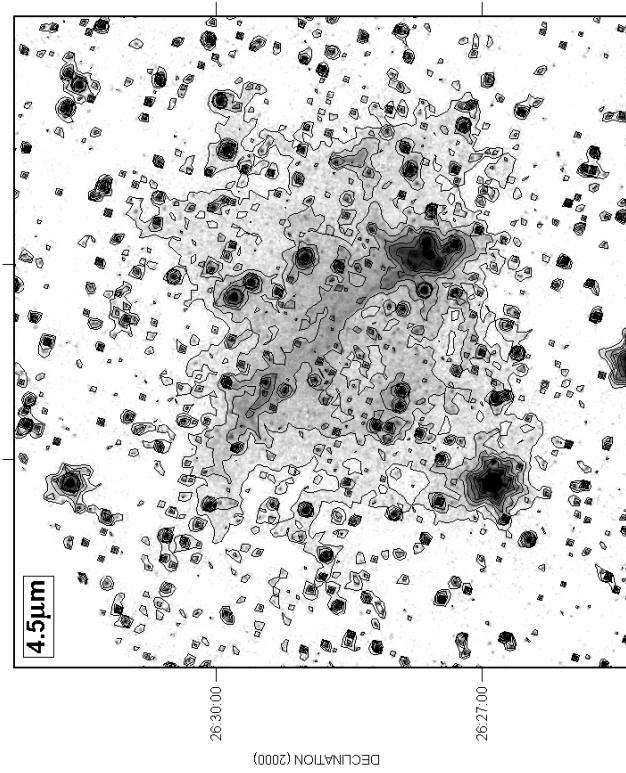
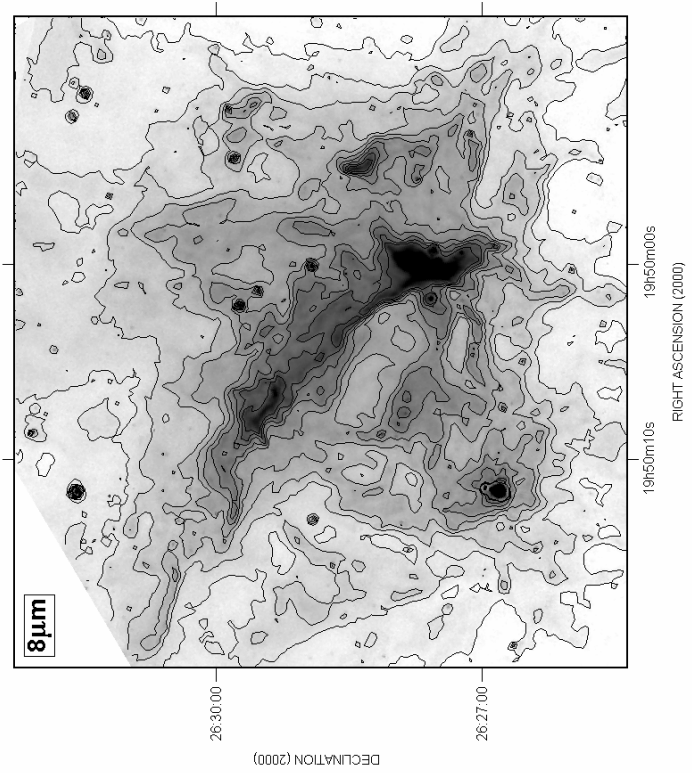
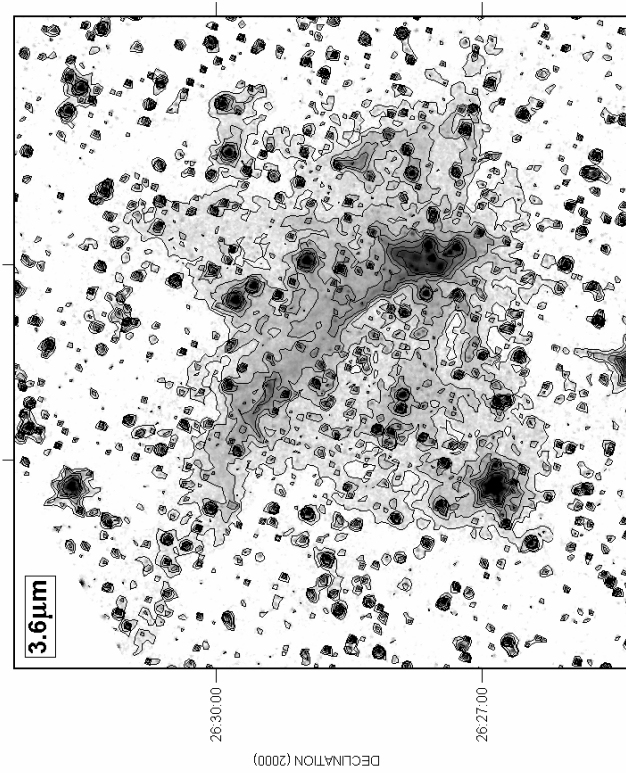
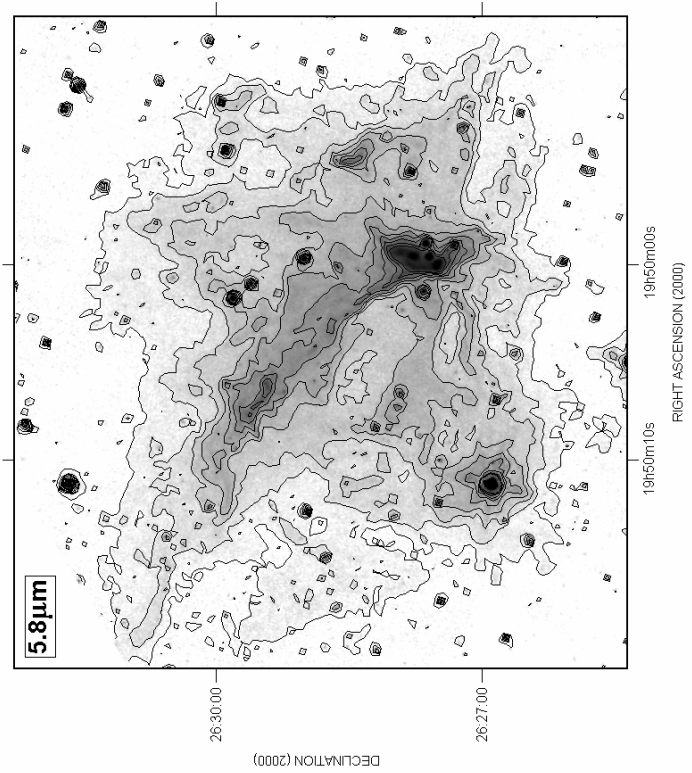

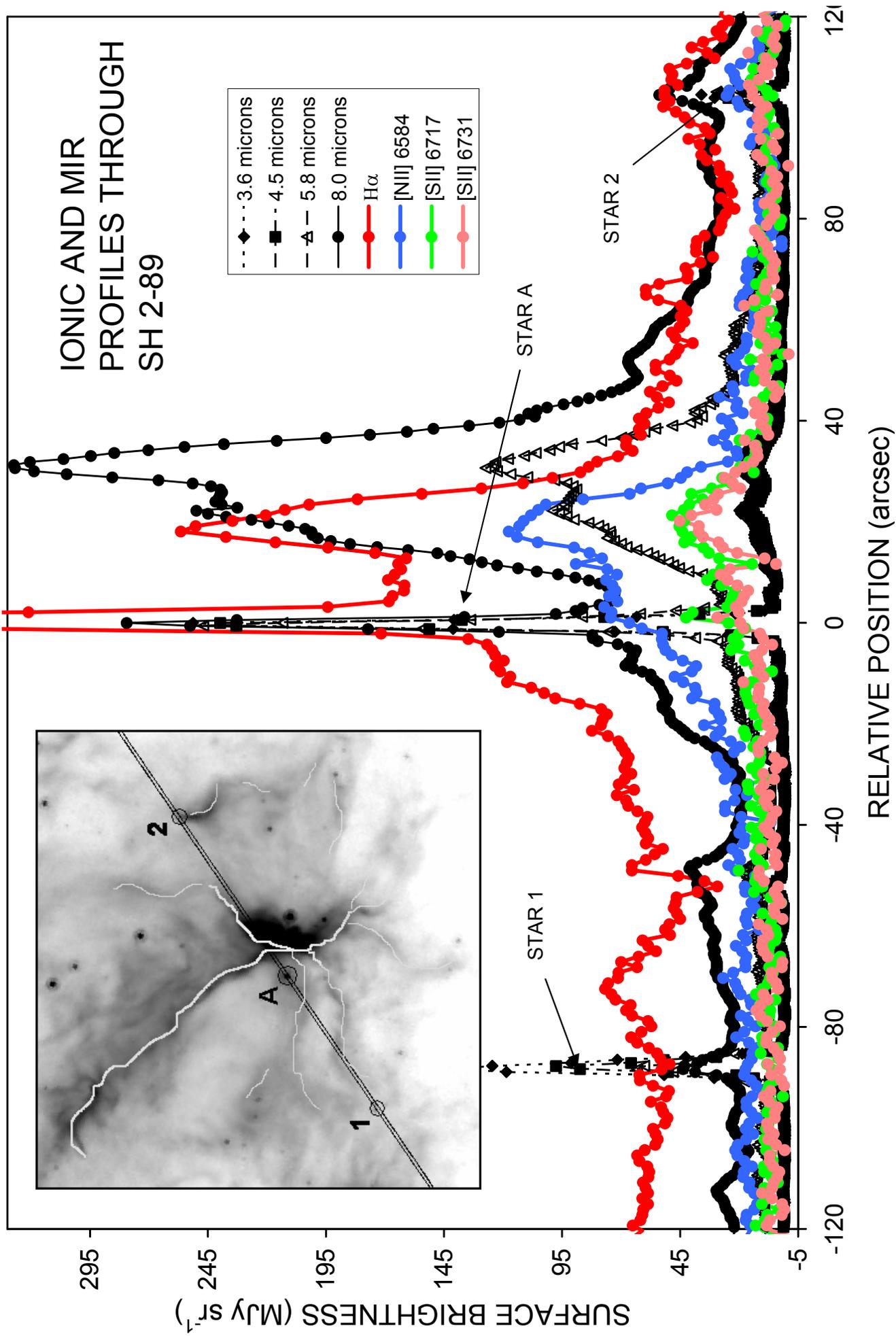

FIGURE 3

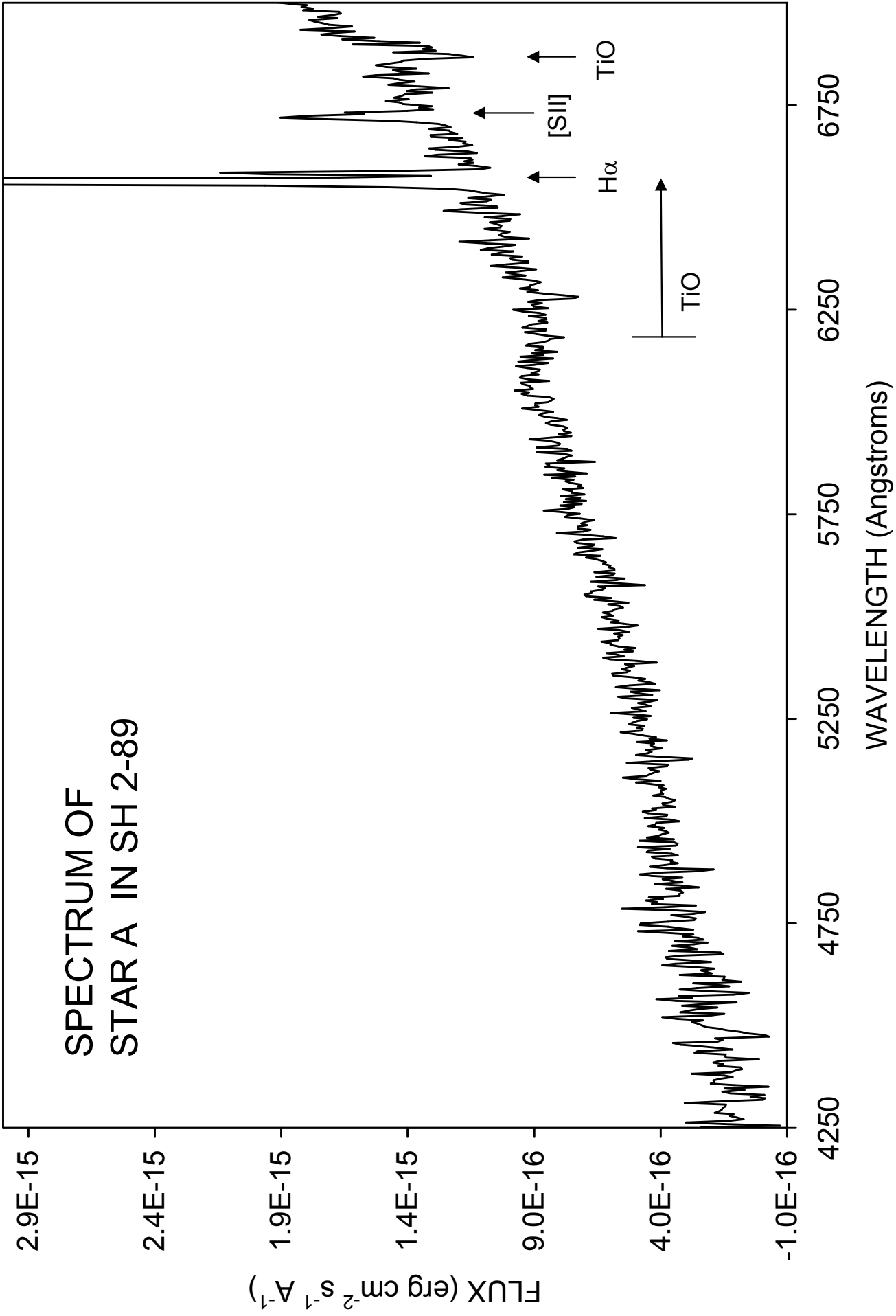

FIGURE 4

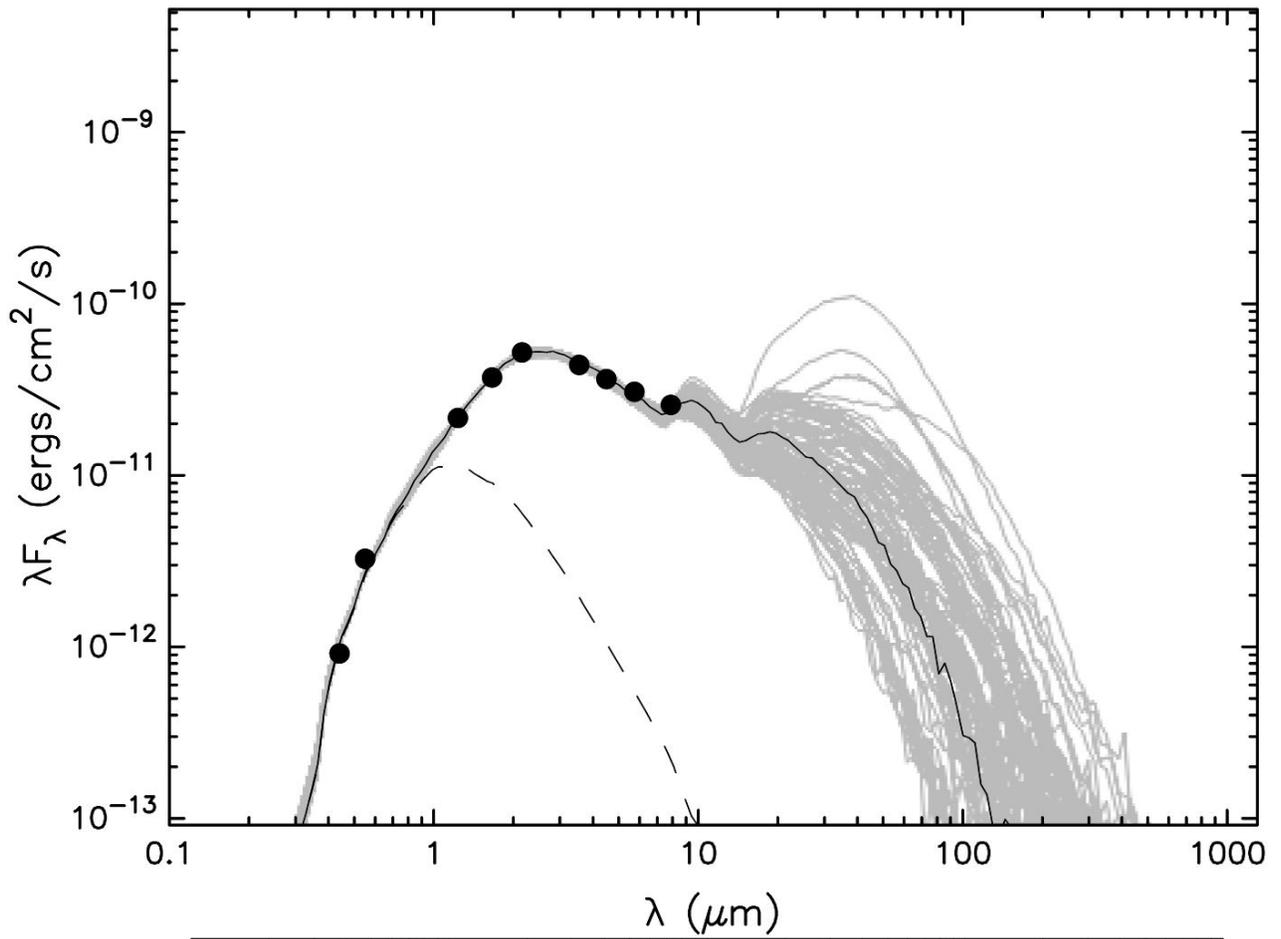
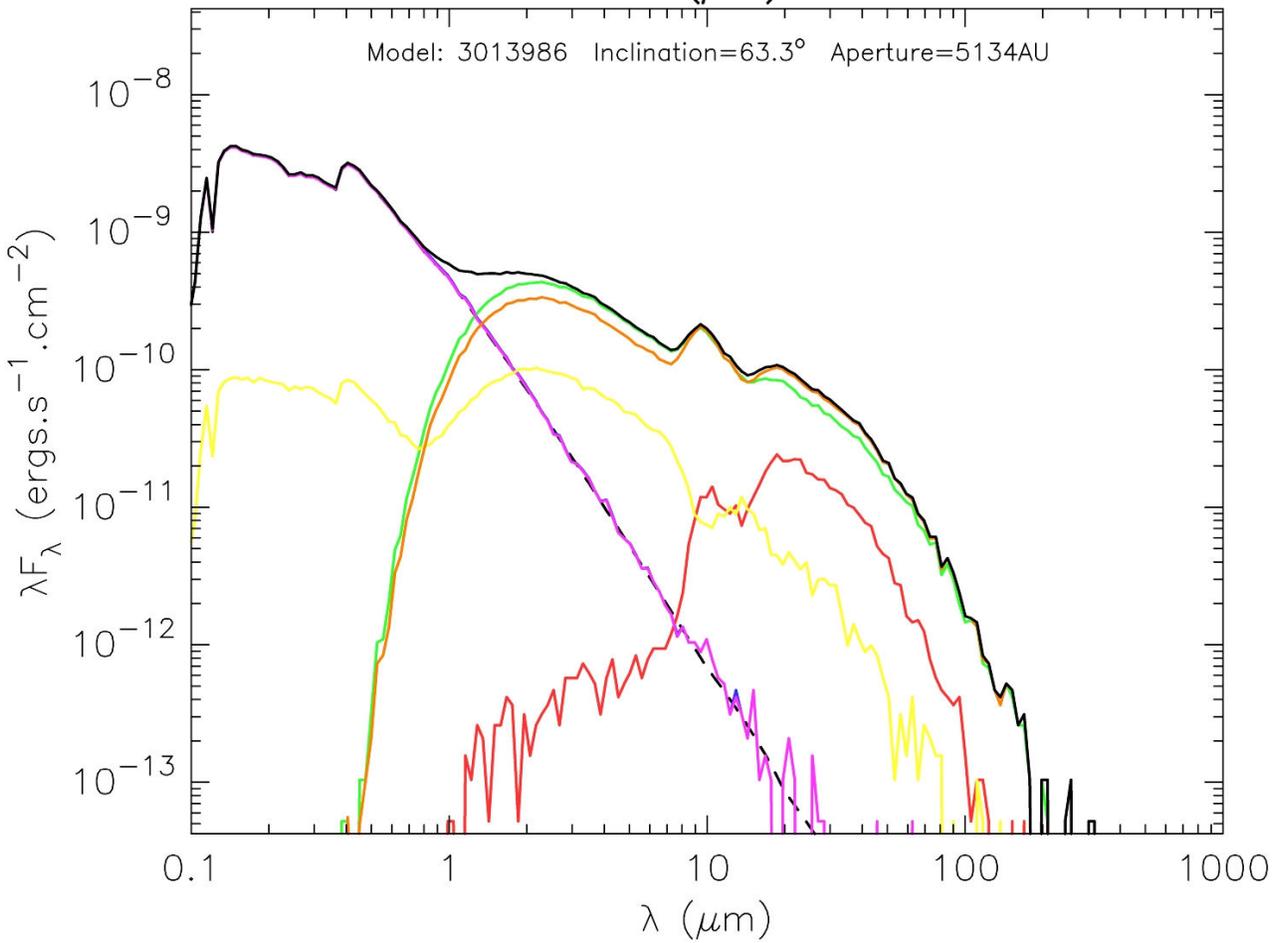

FIGURE 5

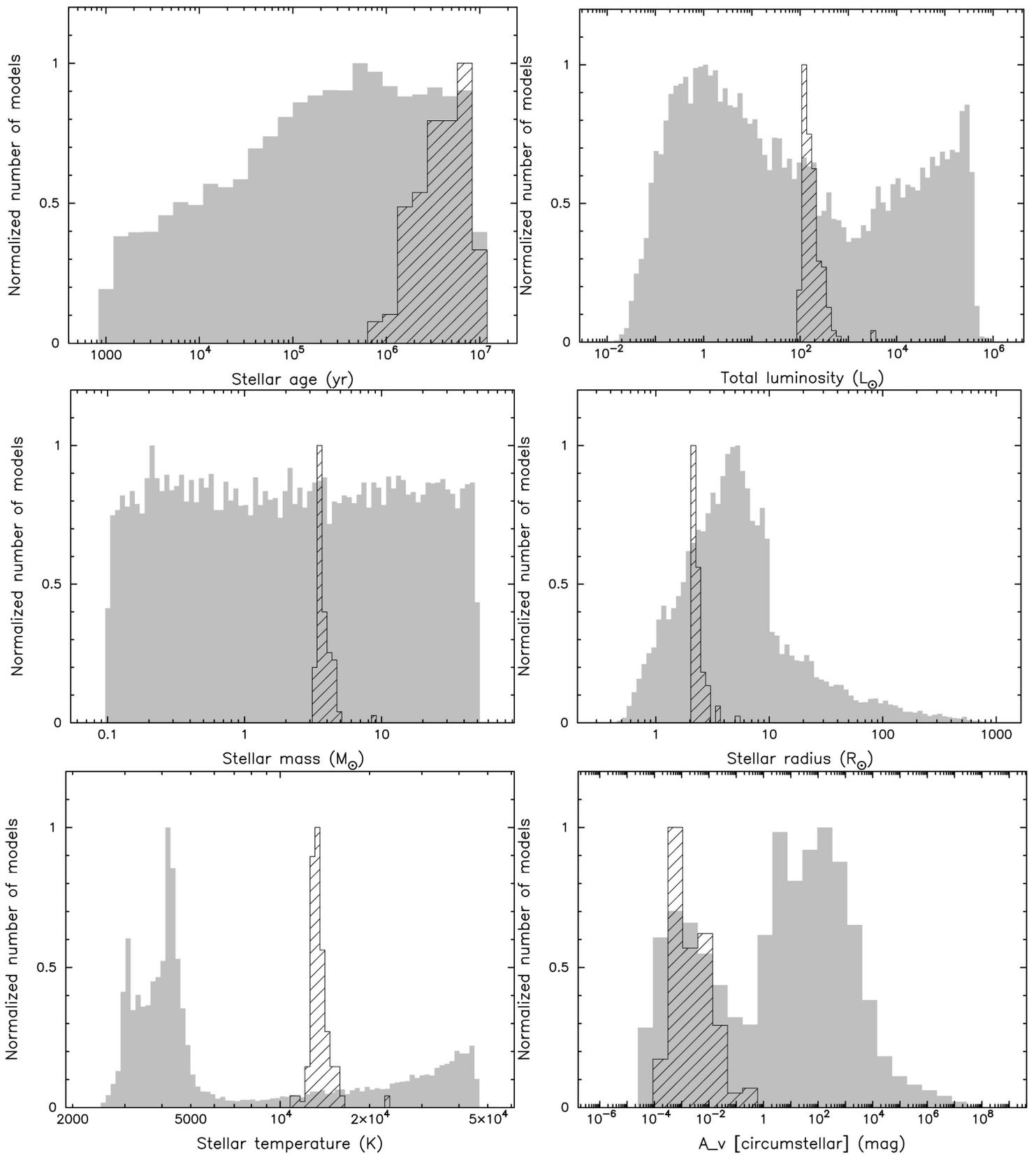

FIGURE 6